\documentclass[twocolumn,aps,floatfix,pra,showpacs]{revtex4}
\usepackage{amssymb,amsfonts,mathrsfs,amsmath}
\usepackage{graphicx,bm}
\usepackage[T1]{fontenc}
\usepackage{pstricks}

\begin{document}

\title{Creation of topological states of a Bose-Einstein condensate in a plaquette}
\author{Tomasz \'Swis{\l}ocki$\,^{1}$, Tomasz Sowi\'nski$\,^{1}$, Miros{\l}aw Brewczyk$\,^2$, and Mariusz Gajda$\,^{1}$}
\affiliation{
\mbox{$^1$ Instytut Fizyki PAN, Al.Lotnik\'ow 32/46, 02-668 Warsaw, Poland}
\mbox{$^2$Wydzia\l {} Fizyki, Uniwersytet w Bia\l ymstoku, ul. Lipowa 41, 15-424 Bia\l ystok, Poland}}

\date{\today}
\begin{abstract}

We study a square plaquette of four optical microtraps containing ultracold $^{87}$Rb atoms in $F=1$ hyperfine state. In a presence of external resonant magnetic field the dipolar interactions  couple initial $m_F=1$ component to other Zeeman sublevels. This process is a generalization of the Einstein-de Haas effect to the case when the external potential has only $C_4$ point-symmetry. We observe that vortex structures appear in the initially empty $m_F=0$ state.  Topological properties of this state are determined by competition between the local axial symmetry of the individual trap and the discrete symmetry of the plaquette. For deep microtraps vortices are localized at individual sites whereas for shallow traps only one discrete vortex appears in the plaquette. States created in these two opposite cases have different topological properties related to $C_4$ point-symmetry. 

%They are quite robust and topological defects created in the Einstein-de Haas effect cannot be changed by subsequent manipulation of the barrier height separating the microtraps.

\end{abstract}
\maketitle

\section{Introduction}
\label{introduction}

Experimental achievement in cooling alkali atoms provide a multiple possibilities to explore Bose-Einstein condensation phenomena. The first Bose-Einstein condensates were created in magnetic traps and allowed for studying atomic condensates properties related to weak contact interactions. Soon after that an optical confinement of ultracold atoms became possible and new systems - spinor condensates - appeared. Here, the ability to simultaneously trap atoms in many magnetic states makes it possible to investigate nontrivial spin dynamics and magnetism in ultracold Bose gases already studied both  
experimentally and theoretically~\cite{Sengstock_1,Chapman,Lewenst,Saito,Gawryluk_1,Swistak}.

It was realized quite early that dipolar interactions between ultracold species might lead to many novel phenomena~\cite{kazik,comment,santos}. The interesting features of dipolar interactions are rooted in their long-range as well as anisotropic character. The Bose-Einstein condensate in $^{52}$Cr atoms, whose magnetic dipole moment equals $ 6 \mu_B$ was obtained after many years of experimental struggles~\cite{pfau_chromium,Beaufils}. This achievement opened a possibility to study a variety of phenomena related to dipolar interactions~\cite{lewenstein_review,ueda_review}. Some authors~\cite{Kurn_prl,Ueda_1,Pu,EdH_GAW,Swistak} noticed however, that magnetic dipole interactions might lead to observable effects even in a gas of rubidium atoms. An example is the Einstein-de Haas effect~\cite{EdH_GAW,EdH_GAW1} (studied also for $^{52}$Cr condensate in~\cite{ueda}). It's main idea is that rubidium atoms trapped initially in $m_F=+1$ component can be efficiently transferred to $m_F=0$ state and then further to $m_F=-1$ Zeeman sublevel. This can happen only if external magnetic field is tuned to a resonant value. Moreover, atoms in $m_F=0$ and $m_F=-1$ states acquire one or two quanta of orbital angular momentum, respectively~\cite{EdH_GAW}.

Ability of making use of interfering laser beams to create optical lattices filled with ultracold bosonic or fermionic atoms opened new possibilities. Atomic physics started to penetrate areas traditionally associated with solid state physics. Observation of the Mott insulator -- superfluid transition in an optical lattice~\cite{Greiner, Stoferle, Spielman} was a very important step towards investigation of strongly correlated systems described by various Hubbard-type Hamiltonians. Since then degenerate gases inside optical lattices are studied very intensively~\cite{Bloch_Dalibard,Lewenstein_Sanpera}.

The progress in experimental techniques allowed to create new types of optical lattices such as superlattices or arrays of coupled plaquettes~\cite{Kruse}. In this article we focus on topological states of the ultracold atomic system in a single plaquette. We study a case of relatively large occupation of the plaquette what allows for the mean field description of the system. Such systems are experimentally accessible~\cite{Pasquiou,Pasquiou_1}.  We concentrate on the topological phases of the macroscopic order parameter. Creation of these dynamical states implicitly utilizes the dipolar interactions tuned by the external magnetic field. External magnetic field is a perfect tool for manipulating the system in the controlled manner. The analysis presented in this article becomes important especially in the context of recently performed experiment~\cite{Pasquiou_1} aiming at the observation of the Einstein-de Haas effect in a chromium condensate in an optical lattice.

If the atoms are located in an array of traps then the Einstein-de Haas effect may exhibit a new interesting aspects. Conservation of the total angular momentum is the essence of the Einstein-de Haas effect. In this article we assume that $^{87}$Rb atoms are placed in a square plaquette of four optical traps. Therefore, the external potential has no axial symmetry and $z$-component of the total angular momentum has not to be conserved. Instead, the $C_4$ point-symmetry comes into a play. The aim of this article is to investigate the role of this symmetry in the dynamics driven by the dipolar interactions. The relative weight of the local axial symmetry of the trap and the discrete symmetry of the plaquette can be controlled by varying the height of the barrier separating the traps.

This paper is organized as follows. In Sec. \ref{description} we describe the system under consideration. In particular, we discuss in detail the properties of dipolar interactions which are essential for the physics reported in the article. Sec. \ref{EdHeffect} presents numerical results regarding the Einstein-de Haas effect for a rubidium condensate in a plaquette whereas Sec. \ref{discrete_local} explains why dipolar resonances sometimes lead to the global discrete vortex state while otherwise to the array of local vortices. In Sec. \ref{stability} we discuss the stability of states obtained via the resonant Einstein-de Haas effect. We end with conclusions in Sec. \ref{conclusions}.

\section{Description of the system}
\label{description}

The single-particle Hamiltonian has the following form:
\begin{equation}
\label{single_particle}
H_0= \int {\rm d}{\bf r} \phantom{1} \hat{\psi}^{\dagger} ({\bf r})\left( \frac{\bf{p}^2}{2m}+V_{tr}({\bf r})-\gamma\, {\bf B} \hat{{\bf F}}\right)\hat{\psi}({\bf r}) \,,
\end{equation}
where $\hat{\psi}({\bf r})=(\hat{\psi}_+({\bf r}), \hat{\psi}_0({\bf r}), \hat{\psi}_-({\bf r}))^T$ is the spinor annihilation operator, $\hat{{\bf F}}=(\hat{F}_x, \hat{F}_y, \hat{F}_z)$ are standard $F=1$ spin matrices, and $\gamma$ is the gyromagnetic coefficient. The external potential of the plaquette $V_{tr}$ is:
\begin{eqnarray}
\label{trap}
V_{tr}({\bf r}) &=& V_0 \left(\cos^2(k_0 x)+\cos^2(k_0 y)\right)  \nonumber \\
&+& \frac{1}{2}m (\omega^2_\bot (x^2+y^2)  +  \omega_z z^2)  \,.
\end{eqnarray} 
The first term describes periodic lattice while the second one corresponds to a relatively weak harmonic potential. The harmonic confinement is essential when the stability of topological states created by resonant Einstein-de Haas effect is discussed (Sec. \ref{stability}). Otherwise (Secs. \ref{EdHeffect} and \ref{discrete_local}), we assume that only axial (i.e., along $z$-axis) part of harmonic trapping is present. The lattice is defined on a square of the $xy$ plane centered at $x=y=0$ and containing four lattice minima which form  a square and are separated by  $d_0=\pi/k_0$. $V_0$ is a potential barrier separating lattice sites. The last term in~(\ref{single_particle}) is responsible for the linear Zeeman shift due to the uniform magnetic field ${\bf B}$ directed along the $z$-axis. In the following we assume that the magnetic field is weak and therefore the quadratic Zeeman effect can be neglected.

The short-range interactions are typically described by a pseudo-potential. In the case of spinor $F=1$ condensate this interaction  can be splited into the spin-independent term proportional to $c_0=4\pi \hbar^2(2a_2+a_0)/(3m)$ and the spin-dependent part proportional to $c_2=4\pi \hbar^2(a_2-a_0)/(3m)$~\cite{HO} where $a_2$ $(a_0)$ is a scattering length of two colliding atoms with total spin $F=2$ $(F=0)$. The Hamiltonian describing the short-range interactions can be written in the form:
\begin{equation}
\label{contact}
H_C=\int {\rm d}{\bf r} \phantom{1} \left(\frac{c_0}{2} : n^2({\bf r}): +\frac{c_2}{2} :{\bf F}^2: \right) ,
\end{equation}
where $n({\bf r})=\sum \psi^{\dagger}_s \psi_s$ is the total atom density, $::$ denotes normal ordering, and square of the total spin operator can be written as: ${\bf F}^2({\bf r})=2\,(F_+({\bf r})F_-({\bf r})+F_-({\bf r})F_+({\bf r}))+F_z({\bf r})F_z({\bf r})$. The raising (lowering) operators of the $z$-component of the spin of the atom at position $\bf r$ are defined as: 
\begin{equation}
\label{F-matrix}
F_{\pm}({\bf r}) = \hat{\psi}^{\dagger}({\bf r})\, \frac{ \hat{F}_x \pm i \hat{F}_y}{2}\, \hat{\psi}({\bf r}),
\end{equation}
while the magnetization density is:
\begin{equation}
\label{magnetization}
F_{z}({\bf r})= {\hat \psi}^{\dagger}({\bf r})\,  \hat{F}_z  \, \hat{\psi}({\bf r})   \,\,.
\end{equation}
Finally, the long-range dipolar Hamiltonian is:
\begin{equation}
\label{H_D_1}
H_{D}=\frac{\gamma^2}{2} \int{\rm d}{\bf r} \int{\rm d}{\bf r}^{\prime} 
:\frac{{\bf F}({\bf r}){\bf F}({\bf r}^{\prime})-3\left[{\bf F}({\bf r}){\bf n}\right]\left[{\bf F}({\bf r}^{\prime}){\bf n}\right]}{|{\bf r}-{\bf r}^\prime|^3}:,
\end{equation}
where ${\bf n}$ is a unit vector in the direction of ${\bf r}-{\bf r}^{\prime}$. Using explicit form of spin-1 matrices the above Hamiltonian can be brought to the form:
\begin{equation}
\label{H_D}
H_{D}=\frac{1}{2}\int {\rm d}{\bf r} \int{\rm d}{\bf r}^{\prime} \frac{\gamma^2}{|{\bf r}-{\bf r}^\prime|^3} :h_D({\bf r},{\bf r}^{\prime}):  \,\,, 
\end{equation}
where $h_D({\bf r},{\bf r}^{\prime})$ has the form:
\begin{eqnarray}
\label{Hdip}
h_{D}&=&A\, \left(F_z({\bf r}^\prime) F_z({\bf r})-F_{+}({\bf r}^\prime)F_{-}({\bf r})   
-F_{-}({\bf r}^\prime)F_{+}({\bf r}) \right) \nonumber \\
&-& 3 \sin^2\ \theta\left(e^{-2i\phi} F_{+}({\bf r}^\prime) F_{+}({\bf r})+e^{2i\phi} F_{-}({\bf r}^\prime) F_{-}({\bf r})\right)  \nonumber \\
&-& 3/2 \sin 2\theta\,  e^{-i\phi}  \left(F_{+}({\bf r}^\prime) F_z({\bf r})+F_z({\bf r}^\prime)F_+({\bf r})\right)   \nonumber \\
&-& 3/2 \sin 2\theta \,  e^{ i\phi} \left(F_z({\bf r}^\prime) F_{-}({\bf r}) + F_{-}({\bf r}^\prime) F_z({\bf r}) \right).  
\end{eqnarray}
$\theta$ and $\phi$ are the spherical angles of the vector ${\bf r}-{\bf r}^\prime$ connecting two interacting atoms and $A=1-3\cos^2\theta$. The form of~(\ref{Hdip}) allows for physical interpretation of all terms.  The first line represents dipolar interactions which do not lead to change of total magnetization of the field: $z$-components of spin of both interacting atoms remain unchanged or the $z$-component of one atom decreases by one while the $z$-component of the second atom increases by one. The second line collects terms describing processes where both interacting atoms simultaneously flip the $z$-axis projection of the spin: both by $+1$ or both by $-1$. Notice that corresponding terms are multiplied by the phase factor $e^{-2i\phi}$ or $e^{2i\phi}$, respectively. This corresponds to change of the projection of the orbital angular momentum of atoms in their center of mass frame by $-2$ or $2$ quanta. The last two lines describe processes in which the spin of one interacting atom is unchanged while the $z$-axis component of the spin of the other atom changes by $+1$ (or by $-1$). This spin flipping term is multiplied by the phase factor $e^{-i\phi}$ (or $e^{i\phi}$) what signifies the change of the $z$-projection of relative orbital angular momentum of interacting atoms by $-1$ (or $+1$ respectively). Evidently, the dipolar interactions conserve the $z$-projection of the total angular momentum of interacting atoms.

\section{Einstein-de Haas effect in a plaquette}
\label{EdHeffect}

We study a system of $N=4\times 10^3$ $^{87}$Rb atoms initially in the polarized $m_F=+1$ state and the ground state of the external potential. Because occupation of this state is macroscopic we describe the system by a spinor wavefunction satisfying the Gross-Pitaevskii (GP) equation. We therefore replace the field operator $\hat{\psi} ({\bf r})$ by the $c$-number spinor wavefunction $\psi ({\bf r})=(\psi_+({\bf r}), \psi_0({\bf r}), \psi_-({\bf r}))^T$. Standard optical lattices for rubidium atoms have the site separation in the submicron regime. In such a geometry and the number of atoms considered here the tunneling is large even for barriers much larger than it is accessible experimentally. However, it is possible to suppress the tunneling by using larger separations between microtraps. Large separations can be reached in various ways, for example by varying the angle between co-propagating laser beams creating a periodic potential.

We solve the spinor GP equation with the nonlocal dipolar potential on a spatial grid $32 \times 32 \times 16$ points. The spatial steps are $dx=dy=0.25\mu$m and $dz=0.5\mu$m, the wavevector is $k_0=0.74/\mu$m, and the axial confinement is given by $\hbar \omega_z/E_R=0.32$, where $E_R=\hbar^2 k^2_0/2m$ is the single-photon recoil energy. Note that at each site, the microtrap potential has a prolate shape due to the tight confinement in the $xy$ plane in comparison with the one along the $z$-direction according of the harmonic trap. Here, we assume that the radial harmonic confinement is absent ($\omega_\bot=0$). We instantly turn on the magnetic field and monitor dynamics of the spinor wavefunction. Transfer of atoms to other Zeeman states strongly depends on the value of the external magnetic field and geometry of the trap~\cite{EdH_GAW}. This transfer is possible only when the Zeeman energy compensates for the excitation energy. Therefore we have to adjust the value of the external magnetic field to the resonant value. For a given initial state there are many magnetic resonances~\cite{Swistak_1}. We focus on the first resonance, corresponding to the smallest value of the magnetic field and the most efficient transfer of atoms to other Zeeman components. We start our simulations with relatively low barrier between plaquette sites, $V_0=10 E_R$. After about $t=4$s, at resonant magnetic field $B=0.075\,$mG, a transfer of atoms to $m_F=0$ component reaches maximum and starts do decrease. We observe a kind of oscillations of population at longer time scale if magnetic field is on. To check the stability of the created dynamically state we switch off the magnetic field at the moment when the transfer is maximal. This way the dipolar interactions are tuned out of the resonance. Also the spin-mixing contact term is small because the number of atoms in $m_F=0$ component is of the order of $100$ only. Therefore further dynamics looks stationary at least at a time scales of few seconds. Population of $m_F=-1$ component remains always very small, however its topological structure proves dipolar character of atoms transfer. Analysis of a one-particle density matrix averaged over $z$-axis~\cite{rev} indicates that all three components of the spinor wavefunction are coherent. In the following discussion we ignore $m_F=-1$ component because of its negligible occupation. It is, however, present in our calculations.

\begin{figure}[!htb]
\begin{center}
\resizebox{3.4in}{1.05in}{\includegraphics{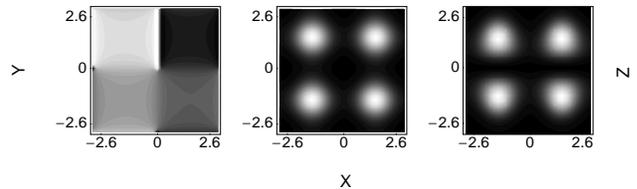}}
\caption[]{Phase (left), the density (middle) in the $xy$ plane at $z=1.3/k_0$ and the density in the $xz$ plane at $y=1.5/k_0$ (right) of $m_F=0$ component at $t=4\,$s for $V_0=10 E_R$ in the case without radial harmonic confinement. Here, $k_0=0.74/\mu$m and the distance between lattice sites equals $d_0=\pi/k_0=4.24\mu$m. Total number of atoms $N=4 \times 10^3$ and the resonant magnetic field $B=0.075\,$mG. Maximal transfer to $m_F=0$ component is equal to $N_0=50$ atoms at $t=4\,$s.}
\label{notrap_rez_1}
\end{center}
\end{figure}

In Fig.~\ref{notrap_rez_1} we present the density distribution and the phase of $m_F=0$ component. Atomic cloud is concentrated around two horizontal $xy$ planes, the one at $z=1.3/k_0$ and the second at $z=-1.3/k_0$. In both planes the density is the same. Discrete symmetry of the plaquette is clearly visible -- almost all atoms are located at centers of the quadrants of the plaquette. The phase of the wavefunction is constant in each quadrant but is not constant over the entire plaquette: it jumps by $\pi/2$ between neighboring quadrants. As a result the phase of the atomic wavefunction winds up by $2\pi$ around the plaquette center. Evidently, the singly quantized discrete vortex in $m_F=0$ component is present. The geometry of the final state proves that the discrete symmetry of the whole plaquette determines the dynamics of the Einstein-de Haas effect. It dominates the local axial symmetry of the individual site in this case.       

\begin{figure}[!htb]
\begin{center}
\resizebox{3.1in}{2.in}{\includegraphics{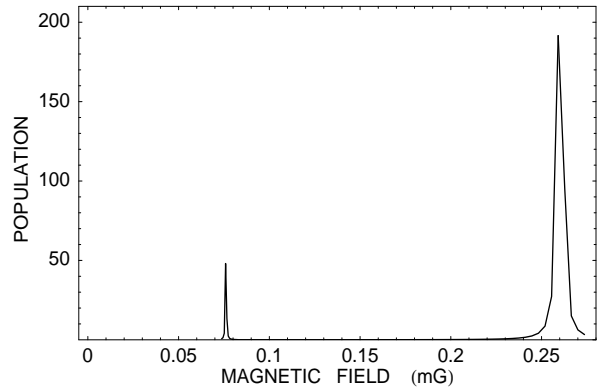}}
\caption[]{Transfer of atoms to $m_F=0$ state as a function of magnetic field for the plaquette with $V_0=10 E_R$. Two resonances are visible. The one for lower magnetic field corresponds to the case when the transfer of atoms is governed by the discrete symmetry of the plaquette (see Fig. \ref{notrap_rez_1}). For the second resonance (for higher magnetic field) the transfer to $m_F=0$ state occurs locally, at individual plaquette sites (see Fig. \ref{notrap_rez_2}).}
\label{N0vB}
\end{center}
\end{figure}

\begin{figure}[!htb]
\begin{center}
\resizebox{3.4in}{1.05in}{\includegraphics{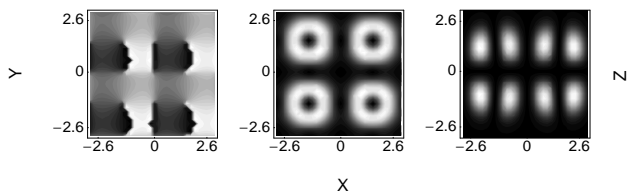}}
\caption[]{Phase (left), the density (middle) in the $xy$ plane at $z=1.3/k_0$ and the density in the $xz$ plane at $y=1.3/k_0$ (right) of $m_F=0$ component at $t=4\,$s for $V_0=10 E_R$ in the case without radial harmonic confinement. Here, $k_0=0.74/\mu$m and the distance between lattice sites equals $d_0=\pi/k_0=4.24\mu$m. Total number of atoms $N=4 \times 10^3$ and the resonant magnetic field $B=0.26\,$mG. Maximal transfer to $m_F=0$ component is equal to $N_0=180$ atoms at $t=0.6\,$s.}
\label{notrap_rez_2}
\end{center}
\end{figure}

However, the topology of the final state might change. It turns out that there exists a resonance which populates the state with local vortices located at the lattice sites. To find such a resonance we still keep the height of the barrier separating the sites at $V_0=10 E_R$ and increase the magnetic field. We observe a significant transfer of atoms to $m_F=0$ state for the magnetic field equal to $B=0.26\,$mG (see Fig. \ref{N0vB}). In  Fig.~\ref{notrap_rez_2} we show the density distribution of $m_F=0$ component at the maximum of transfer. Similarly as in the previous case the final state does not change noticeably when evolved with external magnetic field set to zero. The number of $m_F=0$ atoms in each site is equal to $N_0=180$. We see the array of singly quantized vortices. At each lattice sites the density forms two rings -- the one above $z=0$ plane and the second below this plane. The phase of the corresponding wavefunction winds up by $2\pi$ around every site center. Even more, the phases of individual vortices seem to be correlated because lines where the phase is zero are parallel. Evidently, in the case of this resonance the transfer of atoms to $m_F=0$ component takes place locally, at every lattice site and resembles the Einstein-de Haas effect in axially symmetric harmonic potential. Note that the potential in our case is axially symmetric only close to the minimum at every site of the plaquette. 

\section{Appearance of discrete vortex and local vortices states}
\label{discrete_local}

To get better understanding of our numerical results we will limit our spinor wavefunction to a few relevant states only. At every site localized at $\boldsymbol{r}_i=(x_i,y_i,0)$ where  $i=1,\dots,4$ (for $2 \times 2$ plaquette) we keep in the $m_F=1$ component only the ground state:
\begin{equation}
\langle \boldsymbol{r}|i\rangle_g = {\cal N}_0 {\cal W}_0(x-x_i,y-y_i) \mathrm{e}^{-\beta z^2/4} .
\end{equation}
Two-body dipolar interactions can transfer only one particle from the initial ground state $\langle \boldsymbol{r}|i\rangle_g$ to excited states with $m_F=0$. Simultaneous transfer of both particles is strongly suppressed for large initial state occupation \cite{Swistak_1}. There are two relevant low energy excited states $\langle \boldsymbol{r}|i\rangle_{e_1}$ and $\langle \boldsymbol{r}|i\rangle_{e_2}$ in a given site which can be resonantly populated. The first one is the state with no orbital angular momentum and one $z$-excitation quantum:
\begin{equation}
\label{e1}
\langle \boldsymbol{r}|i\rangle_{e_1} = {\cal N}_1  {\cal W}_0(x-x_i,y-y_i) z \mathrm{e}^{-\beta z^2/4},
\end{equation}
while the second one has one quantum of orbital angular momentum and one $z$-excitation quantum:
\begin{equation}
\label{e2}
\langle \boldsymbol{r}|i\rangle_{e_2} = {\cal N}_2  {\cal W}_1(x-x_i,y-y_i) z \mathrm{e}^{-\beta z^2/4}.
\end{equation}
${\cal W}_0(x,y)$ is the 2D ground Wannier wavefunction with zero orbital angular momentum being a product of 1D ground Wannier states ${\cal W}_0(x,y)=W_0(x) W_0(y)$ while  ${\cal W}_1(x,y)=\left[W_1(x)W_0(y)+iW_0(x)W_1(y)\right]/\sqrt{2}$ is the 2D excited Wannier state with one quantum of orbital angular momentum along the $z$-axis. Here $W_1$  is the first excited 1D Wannier state, $\beta = \hbar\omega_z/E_R$ is the harmonic oscillator energy expressed in units of the recoil energy while ${\cal N}_0$, ${\cal N}_1$, ${\cal N}_2$ are normalization constants. For large occupation of the initial state the processes leading to the chosen excited states strongly dominate others as discussed in \cite{Swistak_1}. We will neglect the $m_F=-1$ component because its occupation is due to the second order perturbation in dipolar interactions. The first excited state (\ref{e1}) could, in principle, be responsible for appearance of discrete vortex in the entire plaquette, while the second one (\ref{e2}) corresponds to the vortex array.

Mean energies of all above states do not depend on the lattice site and they are $E_g$, $E_{e_1}$ and $E_{e_2}$ respectively. All states, i.e. the ground and the two excited states are four-fold degenerate in the $2 \times 2$ plaquette.  Tunneling however couples neighboring sites and partially removes this degeneracy. In the chosen basis the spectrum of the single particle Hamiltonian splits into a three `bands' of four Bloch states:  
\begin{subequations}
\begin{align}
|\!\!\downarrow\rangle_{\alpha} &= \frac{1}{2} \left(|1\rangle_{\alpha} + |2\rangle_{\alpha} + |3\rangle_{\alpha} + |4\rangle_{\alpha}\right), \\
|\!\!\circlearrowleft\rangle_{\alpha} &= \frac{1}{2} \left(|1\rangle_{\alpha} + i|2\rangle_{\alpha} - |3\rangle_{\alpha} -i |4\rangle_{\alpha}\right), \\
|\!\!\circlearrowright\rangle_{\alpha} &= \frac{1}{2} \left(|1\rangle_{\alpha} - i|2\rangle_{\alpha} - |3\rangle_{\alpha} +i |4\rangle_{\alpha}\right), \\
|\!\!\uparrow\rangle_{\alpha} &= \frac{1}{2} \left(|1\rangle_{\alpha} - |2\rangle_{\alpha} + |3\rangle_{\alpha} - |4\rangle_{\alpha}\right), 
\end{align}
\end{subequations}
where  $\alpha=g,e_1,e_2$  and the state vector $|i\rangle_{\alpha}$ corresponds to the wavefunction $\langle \boldsymbol{r}|i\rangle_{\alpha}$.
Energies of the four states within a given band are equal to $\{E_{\alpha}-2J_{\alpha}, E_{\alpha}, E_{\alpha}, E_{\alpha}+2J_{\alpha}\}$, respectively. Tunneling coefficients in the ground band and in the first excited bound are positive and equal $J_g=J_{e_1}$ because all states in these bands have the same spatial profile in the x-y plane.  Therefore the states $|\!\!\downarrow\rangle_{g}$ and $|\!\!\downarrow\rangle_{e_1}$ are the lowest energy states in corresponding bands. 

Tunneling between vortex states in the second excited band is negative $J_{e_2}<0$. Therefore the lowest energy state in this band is anti-ferromagnetic $|\!\!\uparrow\rangle_{e_2}$ with alternating relative phases rather then the ferromagnetic one $|\!\!\downarrow\rangle_{e_2}$. Subspace of eigenenergy $E_{\alpha}$ spanned by vectors \mbox{$|\!\!\circlearrowleft\rangle_{\alpha}$} and $|\!\!\circlearrowright\rangle_{\alpha}$ is two-dimensional, so one can choose any combination of these vectors as the basis. We choose these particular combination since such vectors are not only the eigenstates of the Hamiltonian but also they are the eigenstates belonging to the proper representations of the $C_4$ symmetry group with nonzero eigenvalues $\pm 1$, i.e., they are closely related to the eigenstates of the angular momentum operator in the continuous case. 

In our numerical simulations we prepared the system in the $|\!\!\downarrow\rangle_g$ state. Dipolar interactions can couple this state to the both excited bands. The  strength of this coupling is given by the two-body dipolar matrix element between the initial and final two particle states. In the final state one atom remains in the initial state but the second one occupies one of the states from the excited bands. Direct evaluation of the relevant integrals shows that only two of them strongly (by several orders of magnitude) dominate the others. They correspond to the process when one atom is transferred to the discrete vortex state in the first excited band, i.e., the state in which density profile is almost the same as in the initial state but the phase changes from site to site by $\pi/2$:
\begin{eqnarray}
|\!\!\downarrow\rangle_g \rightarrow |\!\!\circlearrowleft\rangle_{e_1} = \frac{1}{2} \left(|1\rangle_{e_1} + i|2\rangle_{e_1} - |3\rangle_{e_1} -i |4\rangle_{e_1}\right), 
\end{eqnarray}  
or to the state with the array of four vortices:  
\begin{eqnarray}
|\!\!\downarrow\rangle_g \rightarrow |\!\!\downarrow\rangle_{e_2} = \frac{1}{2} \left(|1\rangle_{e_2} + |2\rangle_{e_2} + |3\rangle_{e_2} + |4\rangle_{e_2}\right).  
\end{eqnarray}
The energy difference for the first process is $\delta E_1=E_{e_1}-E_g+2J_{g}$ while for the second one is $\delta E_2=(E_{e_2}-E_g)-2(J_{e_2}-J_g)$. Evidently resonant value of the magnetic field for the first process is smaller then for the second one. Moreover, in the case of the second resonance the relative phases of vortices created at individual lattice sites are the same as relative phases in the initial state. The phase between vortices at different lattice sites is inherited from the initial state. If the initial state is $|\!\!\uparrow\rangle_g$ then atoms are transferred to the analogous state in the second excited band, $|\!\!\uparrow\rangle_{e_2}$. 
This observation agrees very well with our numerical results presented in the previous section. 

Now it is clear that due to the dipolar interactions transfer of angular momentum from the spin part to the spatial part can be realized in two ways: by exciting particles in every site to the second excited band  with array of vortices ($e_2$) or by transferring particles to the first excited band - not changing the local density but changing a relative phases between plaquette sites ($e_1$). 

\begin{figure}
\includegraphics[scale=0.6]{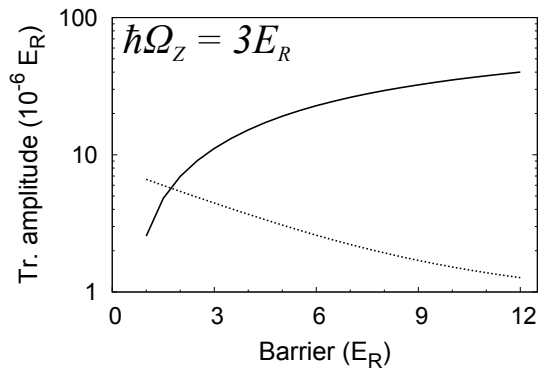}
\caption{Transition amplitudes induced by dipolar interactions for the $2\times 2$ plaquette with $\hbar \omega_z/E_R = 3$. For deeper lattices transition to the four independent vortices located at lattice sites (solid line) always dominates over the transition to the discrete vortex (dotted line). } 
\label{FigWykres}
\end{figure}

Exact values  of these two relevant dipolar transition amplitudes obviously depend on the barrier height and the strength of confinement in $z$-direction (parameter $\beta$). Since for deeply enough lattice the sites become more symmetric, transition to on-site vortices should dominate over the transition to discrete vortex in the plaquette. From the other side, when the lattice is shallow the on-site Wannier states with vortex are quite different from eigenfunctions of `local' (at a given site) orbital angular momentum operators. Moreover the state with discrete vortex, due to a high tunneling, becomes quite almost invariant under rotation around the center of the plaquette by arbitrary angle (up to the phase factor). Therefore we expect that in this case transition to the plaquette vortex will be dominant. We have confirmed all these predictions by numerical calculation of dipolar amplitudes. The results are presented in Fig. \ref{FigWykres}. The height of the barrier for which two considered amplitudes are equal as well as transition amplitudes depend on the parameter $\beta$ characterizing the trapping potential. We should emphasize again that amplitudes which initially do not conserve total angular momentum, i.e., the ones for final states with opposite or without vortex, are numerically many orders smaller than considered ones. 

Presented model is in a very good qualitative agreement with the results of numerical experiment presented here. However, the values of magnetic field when population of the  discrete vortex  dominates population of vortex array state is different. This is because of simplifications we made to get a clear picture of relevant processes. In particular the contact interactions were totally ignored. Their role is to change both energies and spatial spreading of the atomic states what inevitably leads to quantitative differences. 

\section{Stability of topological states}
\label{stability}

Now, we are going to investigate the stability of states populated by resonant Einstein-de Haas effect. For that, we must turn on the radial harmonic confinement since studying the stability of discrete vortex and local vortices states we will first remove the optical potential and next restore it.

We first checked that in the case when the radial harmonic confinement is present together with the periodic potential, we are able to find the same kinds of resonances as in Sec. \ref{EdHeffect}. Indeed, the presence of an extra radial harmonic potential results in a small shift of the resonant magnetic field and a small change in a number of atoms transformed to the $m_F=0$ state as compared to the cases discussed in Sec. \ref{EdHeffect}. For example, for the optical trap with $V_0=10E_R$ and the resonance which populates the singly quantized discrete vortex in $m_F=0$ component we find that the resonance magnetic field equals $B=0.08\,$mG and the number of transferred atoms is $N_0=80$. For further studies we focus on the resonance just mentioned and the other one which leads to significant occupation of vortex states at individual plaquette sites and is found for parameters $V_0=20E_R$, $B_{res}=0.85\,$mG, and $N_0=160$.

\begin{figure}[!bht]
\begin{center}
\resizebox{3.3in}{2.2in}{\includegraphics{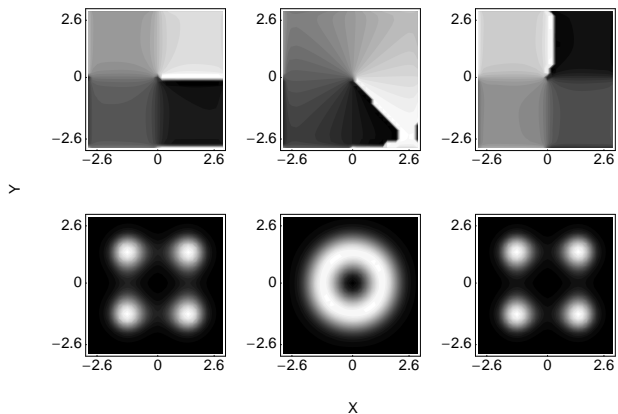}}
\caption[]{Sequence of images of the phase (upper panel) and the density (lower panel) in $xy$ plane ($z=1.3/k_0$) of $m_F=0$ component during lowering and raising of the barrier separating microtraps: $V_0=10 E_R$ (left), $V_0=0 E_R$ (middle), and $V_0=10 E_R$ (right). The barrier was lowered linearly in time during $t=0.5$s. The images in the middle column are shown at $t=0.15$s after the barrier is removed. Then the barrier was raised in $t=1$s and the initial configuration was recovered (see right column). }
\label{movie_1}
\end{center}
\end{figure}

\begin{figure}[!htb]
\begin{center}
\resizebox{3.4in}{2.2in}{\includegraphics{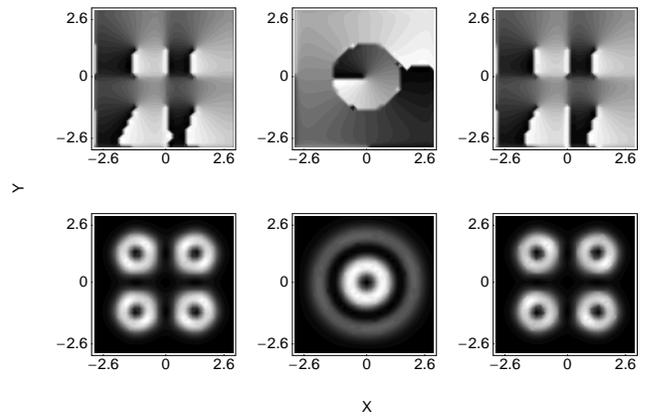}}
\caption[]{Sequence of images as in Fig.~\ref{movie_1} but for $V_0=20 E_R, 0 E_R$ and $20 E_R$ (left, middle, and right column, respectively).  The images in the middle column are shown at $t=0.15$s after the barrier is removed. Then the barrier was raised in $t=1$s and the initial configuration was recovered (see right column). }
\label{movie_2}
\end{center}
\end{figure}

Depending on the barrier separating lattice sites and the value of an external magnetic field the resonant dipolar interactions drive the system towards the states of a different geometry. It is interesting to check if these states are robust against perturbation of the trapping potential of the lattice. To this end we ramp adiabatically the lattice potential down to zero and then we rise it back to the initial value for both studied cases. The sequence of pictures in Figs.~\ref{movie_1},~\ref{movie_2} shows how the density and the phase of $m_F=0$ component change while the microtrap barrier is first lowered and then raised back. The lowering process lasts for $t=0.5\,$s. The middle columns show the density and the phase at $0.15\,$s after the barrier height reached zero. We see that in both cases atoms form a vortex around the center of the plaquette.  In the case of high initial barrier an additional ring of relatively low density surrounding the first one can be seen, Fig.~\ref{movie_2}. However, after raising back the barrier both initial configurations are recovered (see the right columns of Figs.~\ref{movie_1},~\ref{movie_2}) what indicates that the initial states are essentially different and one cannot switch between these states after they are created.

\begin{figure}[!htb]
\begin{center}
\resizebox{3.4in}{1.6in}{\includegraphics{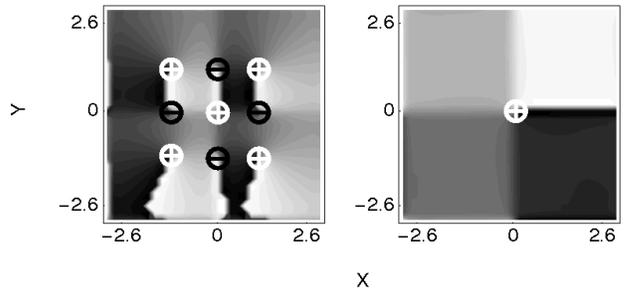}}
\caption[]{Phase portraits of $m_F=0$ component obtained at $V_0=20 E_R$ (left panel) and $V_0=10 E_R$ (right panel). Notice a singly quantized discrete vortex of positive charge corresponding to $V_0=10 E_R$ case. For $V_0=20 E_R$ in addition to four singly quantized vortices of positive topological charge located at the center of each quadrant  where the atomic density is large, there are five more vortices. One of them is located at the plaquette center and the other four  have opposite topological charge and are located at the bonds connecting lattice minima. These additional five vortices are hidden in the region of a  very low atomic density. The vortices and their charges are  marked by white and black circles.  }
\label{phase_12}
\end{center}
\end{figure}

To understand this difference let us look once more into the phases of both states as they were created via dipolar interactions, Fig.~\ref{phase_12}. As the discrete vortex, obtained for lower barrier, $V_0=10 E_R$, is the only singularity of the phase of the $m_F=0$ component (right panel), much richer structure can be observed for the array of vortices obtained at $V_0=20 E_R$ (left panel). In addition to four singly quantized vortices of positive topological charge located at the center of each minimum there is one more vortex of the same charge located at the plaquette center and four more vortices of opposite topological charge. These vortices and their charges are  marked by white and black circles in Fig.~\ref{phase_12}. The additional five vortices are hidden in the region of a  very low atomic density.  The number of vortices strongly differentiates between these states. The vortex structures are stable against perturbation of the lattice potential therefore it is not possible to switch between them by changing the intertrap barrier. In Fig. ~\ref{movie_2} only one vortex is clearly seen. Other vortices are located at regions of vanishing density. However, numerical evaluation of circulation around every point of the grid allows to detect them.  The system recovers the initial symmetry even after the barrier was removed. This memory effect is due to different number of vortices in both studied states. These two states belong to particular representation of the $C_4$ point-symmetry group. This representation is characterized by the presence of singly quantized vortex at the center of the plaquette. It is worth to note at this point that an extensive discussion of solutions of the GP equation belonging to different representations of $C_n$ point-symmetry group can be found in Ref.~\cite{Ferrando}.

We would like to stress that our findings might be important for the observation of the Einstein-de Haas effect in optical lattices. Some experimental effort has been already taken in this direction~\cite{Pasquiou_1}. Our results show that the properties of the rotating states created via the Einstein-de Haas effect strongly depend on the geometry of the trapping potential and the value of the external magnetic field.

\section{Conclusions}
\label{conclusions}

In conclusion, we studied formation of topological states via resonant dipolar interactions. We paid a particular attention to competition between the local axial symmetry of the individual microtrap and global discrete symmetry of the entire plaquette. We showed that the role of these two symmetries can vary depending on the height of the intertrap barrier. For shallow enough plaquettes when the mean field wavefunction penetrates the whole plaquette a single discrete vortex is created. In the opposite case of low tunneling the transfer of atoms has a local character and axial symmetry of individual site allows for formation of the array of vortices located at each plaquette site. However, for a wide range of intermediate lattice's depths both kinds of topological states can be populated depending on the value of applied magnetic field. These topological states are created dynamically with the help of resonant Einstein-de Haas effect. Moreover, the vortex structures present in both states make these states very stable against perturbation of the lattice potential. Even when the discrete symmetry of the plaquette is destroyed by removing the lattice potential, the system comes back to the initial configuration after the lattice potential is raised back.

{\bf Acknowledgment}
The work was supported by the UE project NAME-QUAM and Polish Ministry for Science and Education for 2009-2011.

\end{document}